# An Explanation of the Vaughan – Preston Gap

Peter Foukal; Nahant, MA 01908, USA. pvfoukal@comcast.net

### Abstract

A plot of the calcium emission versus color of late type stars exhibits a reduced population or gap at intermediate activity, somewhat higher than that of the Sun. We suggest that this gap (Vaughan & Preston 1980) may result from a reduced area of plages relative to spots, as observed at the highest levels of solar activity (Foukal 1998a). This reduced plage area weakens the Ca II emission and depletes the number of stars of intermediate Ca HK index. We propose that, in the most active stars, the reduction in *relative* plage area is offset by the increased filling factor of photospheric magnetic fields. So the Vaughan – Preston gap might simply be a consequence of a gradual shift with age of the stellar dynamo towards production of higher spatial frequencies.

### Introduction

Plots of the calcium emission index $R'_{HK}$ against B-V (e.g. Radick et al. 2017) exhibit a reduced population of stars of intermediate activity level, somewhat above that of the Sun (Figure 1). Much younger and more active stars appear more frequent. This population reduction or gap was first pointed out by Vaughan & Preston (1980).

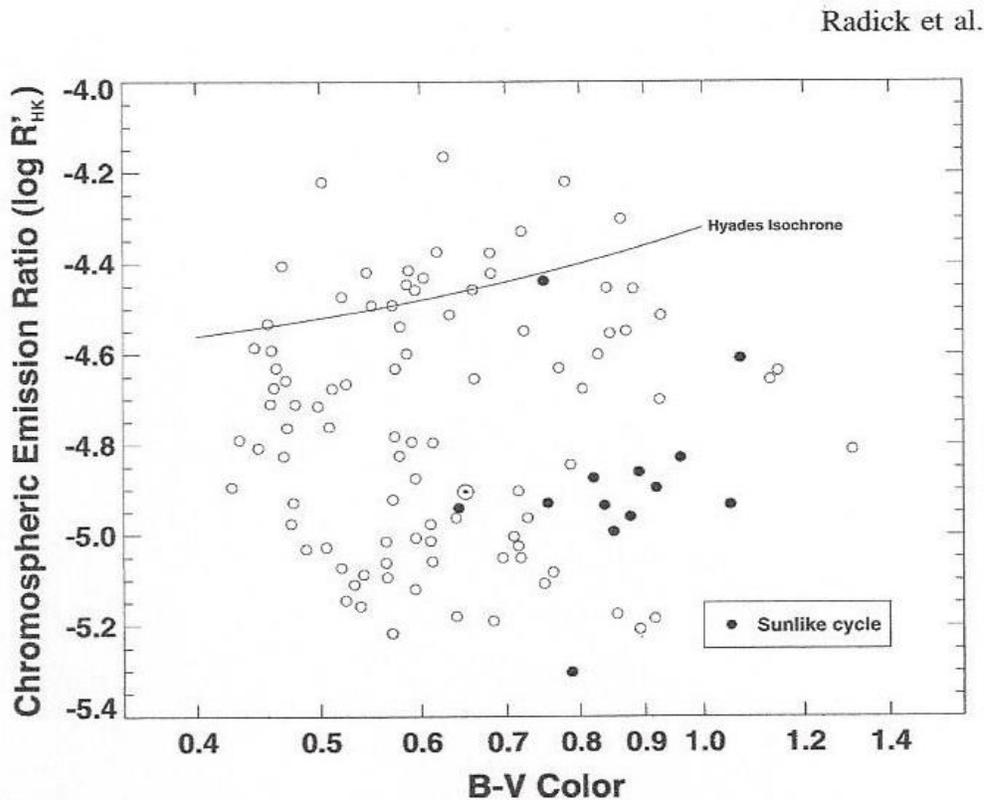

Figure 1. A plot of the chromospheric emission ratio $R'_{HK}$ versus color index B – V (from Radick et al. 2017). The Vaughan – Preston gap refers to the depletion of stars of B – V roughly > 0.6 and - 4.7 < $R'_{HK}$ < - 4.9.

Explanations invoking two different modes of dynamo operation above and below the gap, or of a transition in dynamo behavior at a critical Rossby number, have been advanced (e.g. Brandenburg et al. 1998; Tobias 1997; see review by Giampapa 2016). We suggest here that the gap is a simple consequence of the shift from large spot magnetic flux tubes to smaller facular and plage flux tubes as a star's dynamo weakens with age.

This is the same shift that is now recognized to cause the transition from spot - dominated luminosity variation of young stars to facula - dominated variation in older objects (Foukal 1994, Lockwood, Skiff & Radick 1997). Our explanation here of the gap is based on an extension of this same shift, seen in faculae, to a corresponding shift observed in the statistics on the overlying chromospheric plages.

1. **The decrease in area of plages relative to spots with increasing activity**

The areas of plages and spots, $A_p$ and $A_s$, are linearly related at low to moderate solar activity levels (Foukal 1993, Chapman, Dobias, & Arias 2011). But their ratio, $A_p/A_s$, decreases in binned daily data for the largest observed solar cycles 18 and 19 (Foukal 1998a). This reduced slope at highest activity is illustrated in Figure 2. Although cycle 19 was the largest by sunspot number, the decrease is most evident in cycle 18, associated with the appearance in 1947 of the largest spots in the 102 - year RGO record. Here, $A_s$ is the time series of daily sunspot areas measured by the Royal Greenwich Observatory (RGO) between 1874 – 1976 (Hohenkerk, Ladley & Rudd 1967). $A_p$ are daily Ca K plage areas measured from digitized Mt Wilson Observatory spectroheliograms obtained between 1916 – 1984, (Foukal 1998b).

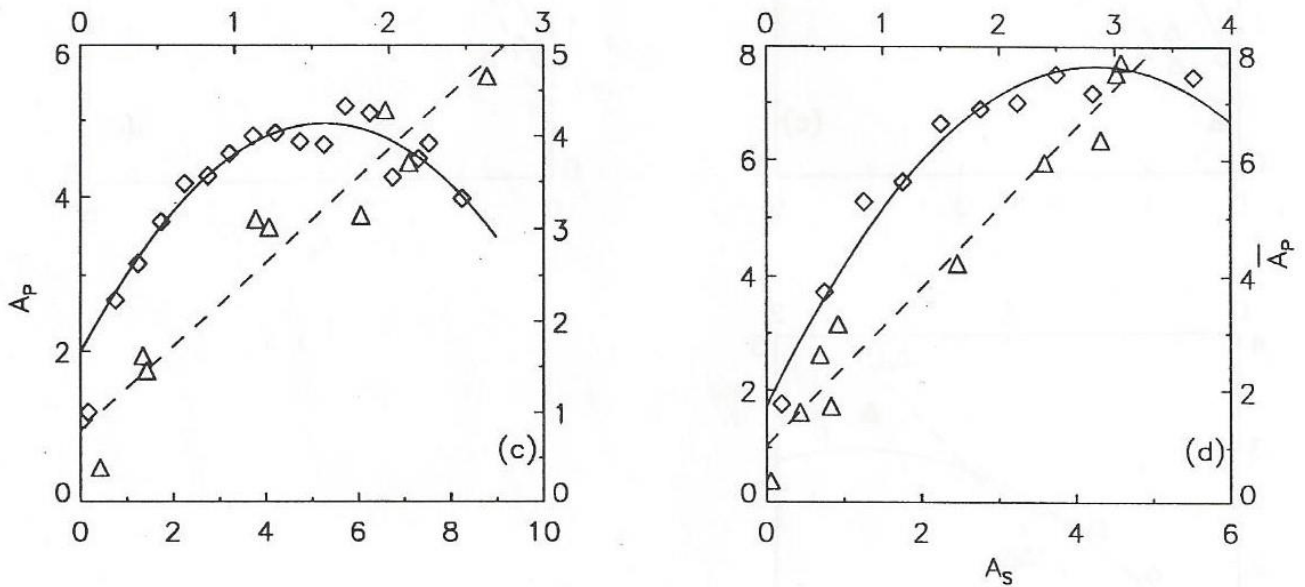

Figure 2. Plots of $A_p$ versus $A_s$ for cycles 18 (left) and 19 (right). Triangles denote annual averages and diamonds represent binned daily data. The abscissa scales at the bottom refer to the binned data; those at the top to the annual averages. Ordinate scales on the left and right refer to binned and annual values of $A_p$ respectively. $A_p$ and $A_s$ are expressed as percent and tenths of a percent of a hemisphere, respectively. From Foukal (1998a).

The decrease seems to be caused by a shift in the spatial power spectrum of stellar magnetic fields towards larger flux tubes and lower spatial wavenumbers as activity increases (Foukal 1993, 1994, 1998a). Such a shift appears consistent with dynamo calculations (e.g. Tobias, Cattaneo & Brummel 2008). This interpretation is supported by a decrease in the area ratio $A_f/A_s$ of white light faculae relative to spots that becomes evident already at moderate solar activity levels (Foukal 1993, 1998a).

The faculae coincide with the very smallest photospheric magnetic flux tubes, so the shift is more evident than in $A_p/A_s$, because $A_p$ is measured from narrow band (0.02 nm) MWO spectroheliograms in which not only the plages overlying the photospheric faculae, but also those overlying the larger pore (and many spot) flux tubes, appear bright in Ca K. So $A_p$ includes not only the areas of faculae but also the areas of larger flux tubes, somewhat masking the more evident shift towards lower spatial frequencies seen in $A_f$ alone.

2. **The reduction of Ca II emission in the gap stars**

Spots emit more weakly than plages in the relatively broad 0.1 nm Ca II passband used in the stellar observations exhibiting the gap. So somewhat more active stars exhibiting lower $A_p/A_s$ than the Sun are expected to emit weaker Ca II relative to their magnetic activity level, than they would if $A_p/A_s$ were independent of activity. Such stars would lie around or below the Sun in a $R'_{HK}$ vs B-V plot, so they would be missing from the gap.

To evaluate this explanation we need to determine by how much the Ca II emission of stars that *should be* in the gap region would need to be reduced to place them at or below the log $R'_{HK}$ ~ - 4.9 level of the Sun. From Figure 1 we see that the displacement in log $R'_{HK}$ is only by roughly 0.1 – 0.2. So the stars of solar B – V that should be in the gap have an average activity level corresponding to Ca II emission strength roughly 25 -50% higher than that of the Sun.

The solar $R'_{HK}$ plotted in Figure 1 was estimated by Radick et al. (2017) from observations made between 1992 – 2016. This period spans the moderate - sized cycle 22 and the two small cycles 23 and 24. The mean activity level during this period is roughly half of that seen in cycles 18 and 19. So the stars of similar B – V to the Sun that are missing in the gap might be expected to have magnetic activity levels similar to the Sun in those two largest observed cycles.

From Figure 2 we see that, in cycle 18, the plage area at maximum spot area is reduced by about 25% below its maximum value reached at a lower $A_s$. Its reduction below the value that a star of its activity *should* have if its Ca II emission were linearly extrapolated from its slope at moderate activity levels is much greater, about a factor of two. The reduction from the extrapolated value in cycle 19 is similar. So it seems plausible that the depletion of stars in the gap is caused by the reduction in plage area relative to that expected if the ratio $A_p/A_s$ remained constant with stellar age.

The stars missing from the gap should be recognizable because they would be expected to exhibit a *negative* correlation of luminosity and $R'_{HK}$. Such a negative correlation would otherwise be puzzling in stars located around or below the Sun since stars of Sun – like or weaker activity level tend to exhibit *positive* correlation over year – to – year time scales.

Several such stars exhibiting negative correlations are found in the Radick et al. (2017) study (see their Figure 14), seemingly in agreement with our hypothesis. But the significance of the correlation sign is low for all besides HD 10307. Also, close examination shows some to be spectroscopic binaries (Egeland

2018). In addition, metallicity and inclination (e.g. Radick et al 1998; Shapiro, Solanki, Krivova et al. 2014) may play a role. More photometry of such stars is required to test this hypothesis.

3. **Restoring the strong Ca II emission in the youngest stars**

The youngest and most active stars of similar B – V to the Sun shown in Figure 1 lie between 0.2 and 0.6 higher in log R'$_{HK}$ than the Sun so they are as much as 4 times brighter in Ca II. Our explanation requires that this brightness be generated despite the decrease in A$_p$/A$_s$. Little is known with confidence about plage brightness and area filling factor on Sun - like stars. But even if plage brightness is taken as constant with activity level (Worden, White & Woods 1998) we require only an increase in filling factor by a factor of about 4 relative to the value representative of the moderate activity in cycles 22-24 to offset the decreased A$_p$/A$_s$ and produce the required Ca II brightness.

The average plage area measured in cycles of such amplitude is approximately 2-3% of the hemisphere (e.g. Foukal 1998b). Somewhat larger values are obtained if enhanced network (which roughly doubles the bright contribution to 11 - year solar total irradiance variation from active region plages (Foukal & Lean 1988)) is included. So, a filling factor greater by a factor 4 is not inconsistent with values around 10% measured on young Sun - like stars (e.g. Saar & Judge 2017).

4. **Discussion and Conclusions**

A decrease of A$_f$/A$_s$ has been measured on the Sun (Foukal 1993) and was subsequently inferred from luminosity variations observed on other late type stars (Lockwood, Skiff & Radick 1997). The accompanying decrease of A$_p$/A$_s$ (Foukal 1998a) may provide a relatively straightforward explanation of the Vaughan – Preston gap. The underlying shift of photospheric fields towards smaller flux tubes does not seem to be caused by a change in the diffusion rate of photospheric fields (Foukal 1998a); its behavior agrees better with change in the source function of erupting magnetic flux, so with change in the dynamo itself. Such a shift may be a property of weakening dynamo action (e.g. Tobias, Cattaneo, & Brummel 2008). To test this explanation further, better statistics are required on stars of R'$_{HK}$ lower than the solar value that exhibit positive correlation of variations in luminosity versus Ca II index variation on year- to - year time scales.

Our findings seem consistent with recent inference from stellar rotation studies of a possible shift towards higher spatial frequencies in the fields of mature late type stars (Metcalfe, Egeland and Van Saders 2016). But the shift inferred in that study refers to the harmonic structure of the global field, whereas the shift discussed here refers to a change in size of the photospheric magnetic flux tubes. A connection between the two, if it exists, requires further study. Metcalfe, Egeland & Van Saders (2016) suggested that their findings might explain the Vaughan – Preston gap, but this suggestion has since been withdrawn (Metcalfe 2018).

Still, their suggestion that the Sun is in a unique transitional state appears to be supported by evidence that it is alone in the sample of 72 Sun – like stars studied by Radick et al. (2017) in switching, during recent activity cycles, between positive and negative correlation of luminosity and activity level (Foukal 2015).